# Transport and Magnetic Studies on the Spin State Transition of $Pr_{1-x}Ca_xCoO_3$ up to High Pressure

Toshiaki Fujita, Takeshi Miyashita, Yukio Yasui, Yoshiaki Kobayashi and Masatoshi Sato

*Department of Physics, Nagoya University, Furo-cho, Chikusa-ku, Nagoya 464-8602*

Eiji Nishibori and Makoto Sakata

*Department of Applied Physics, Nagoya University, Furo-cho, Chikusa-ku, Nagoya 464-8603*

Yutaka Shimojo, Naoki Igawa, Yoshinobu Ishii and Kazuhisa Kakurai

*Japan Atomic Energy Research Institute, Tokai, Ibaraki 319-1195*

Takafumi Adachi, Yasuo Ohishi and Masaki Takata

*Japan Synchrotron Radiation Research Institute,1-1-1 Kouto Mikazuki-cho Sayo-gun Hyogo 679-5198*

Transport and magnetic measurements and structural and NMR studies have been carried out on $(Pr_{1-y}R'_y)_{1-x}A_xCoO_3$ {R'= (rare earth elements and Y); A=(Ca, Ba and Sr) } at ambient pressure or under high pressure. The system exhibits a phase transition from a nearly metallic to an insulating state with decreasing temperature $T$, where the low spin (LS) state of $Co^{3+}$ is suddenly stabilized. For $y=0$, we have constructed a $T$-$x$ phase diagram at various values of the external pressure $p$. It shows that the $(T, x)$ region of the low temperature phase, which is confined to a very narrow region around $x$=0.5 at ambient pressure, expands as $p$ increases, suggesting that the transition is not due to an order-disorder type one. For the occurrence of the transition, both the Pr and Ca atoms seem to be necessary. The intimate relationship between the local structure around the Co ions and the electronic (or spin) state of $Co^{3+}$ ions is discussed: For the smaller unit cell volume or the smaller volume of the $CoO_6$ octahedra and for the larger tilting angle of the octahedra, the temperature of the transition becomes higher. The role of the carriers introduced by the doping of the A atoms, is also discussed. By analyzing the data of $^{59}Co$-NMR spectra and magnetic susceptibilities of $Pr_{1-x}Ca_xCoO_3$ the energy separations among the different spin states of $Co^{3+}$ and $Co^{4+}$ are roughly estimated.

corresponding author : M. Sato (e-mail address:e43247a@nucc.cc.nagoya-u.ac.jp)

## 1. Introduction

Spin state change is one of characteristics of Co oxides, which attracts much interest. In perovskite system $RCoO_3$ (R=Y and various rare earth elements) with three dimensional network of corner-sharing $CoO_6$ octahedra, for example, often exhibits a spin state change from the low spin (LS; spin $s=0$; $t_{2g}^6$) ground state to the intermediate spin (IS; $s=1$; $t_{2g}^5 e_g^1$) or the high spin (HS; $s=2$; $t_{2g}^4 e_g^2$) state with increasing temperature $T$.[1-7] The existence of this spin state change indicates that the energy difference $\delta E$ between these states is rather small. This $\delta E$ value increases with decreasing ionic radius $r$ of $R^{3+}$ through the increase of the crystal field splitting $\Delta_c$ induced by the volume contraction of the $CoO_6$ octahedra. However, it may not be very simple, in general, to understand the detailed change of $\delta E$ of various Co oxides only by considering $\Delta_c$ (and the Hund coupling $J_H$), because the Co-Co transfer energy also depends on $r$ through the changes of Co-O bond length and Co-O-Co bond angle and seems to affect the $\delta E$ value.

In $R_{1-x}A_xCoO_3$ with divalent A (A= Ba, Sr and Ca) atoms, effects of the hole-doping should also be considered besides the effects stated above.[8-15] All these effects seem to change $\delta E$ and the distribution of the electrons in the $3d$ orbitals. Then, to control the physical properties of the system, we have to properly choose the species of R and A atoms and the A atom concentration $x$. External pressure gives another parameter to control the physical properties. Then, it is interesting to search new physical behaviors by choosing proper set of these parameters. The newly found superconductivity of $Na_zCoO_2 \cdot yH_2O$ ($z$~0.3 and $y$~1.3), two-dimensional layers of which consist of $CoO_6$ octahedra, too, may be an example of such unexpected states.[16-19]

We have reported in ref. 8, results of the transport and magnetic measurements on samples of $R_{1-x}A_xCoO_3$ prepared at temperatures lower than 1200 °C and shown that there are rather significant difference of the physical properties between Ba- and Sr-doped systems and Ca-doped one for R=Pr and Nd: For A=Ca with relatively smaller ionic radius, the system in the paramagnetic $T$ region does not become metallic with increasing concentration $x$, while for A=Ba and Sr with relatively large ionic radii, the systems become metallic even in the paramagnetic $T$ region and rather large ferromagnetic moments appear with increasing $x$ at low temperatures. (The actual values of the concentration $x$ of the previous samples are different from the nominal ones, on which we present details later.) The result indicates the importance of the average ionic radius of $R_{1-x}A_x$.

For samples of $Pr_{1-x}Ca_xCoO_3$ prepared at 1200°C, Tsubouchi *et al.* have found a transition at ambient pressure in the very narrow region of $x$~0.5,[15] from the high temperature conducting phase, (which is nearly metallic but in the strict sense it seems not to be metallic, because it does not have the metallic $T$ dependence of the resistivity ρ) to the low temperature less conducting (insulating) phase and proposed that the transition is accompanied by the IS →LS change of $Co^{3+}$ ions with decreasing $T$. We have independently found a transition similar to that observed by Tsubouchi *et al*. in samples of $Pr_{1-x}Ca_xCoO_3$ prepared at 1000 °C in the wide $x$ region under the condition of high pressure $p > 5$ kbar.[8] We have also pointed out that the transition seems to take place only for (R,A)=(Pr,Ca). Although the transition was not observed at ambient pressure even for samples with the nominal concentration $x$=0.5, the behaviors of the resistivity and the magnetic properties at the transition seemed to be very similar to those of the one found by Tsubouchi *et al*. Then, we have carried out further studies by various methods, for samples prepared by changing the sample preparation condition, to clarify details of the transition(s) and to extract more information what factors determine the electronic state of the system. In the studies, we have found, by a combined use of the X-ray and neutron Rietveld analyses, that for samples prepared at relatively low temperatures (~1000 °C), the real values of $x$ are smaller than the nominal value. Here, results of the present studies carried out for various kinds of samples are presented by using the actual values of $x$.

## 2. Experiments

Mixtures of $Pr_6O_{11}$, $CaCO_3$ and $CoC_2O_4 \cdot 2H_2O$ with proper nominal molar ratios were ground, and pressed into pellets. The pellets with $x$=0.5 were sintered at 1000 °C, 1100 °C, 1150 °C and 1200 °C and the pellets with $x$=0, 0.1, 0.2, 0.3 and 0.4 were sintered at 1200 °C, for 12 h under flowing oxygen and cooled at a rate of 100 K/h. Samples thus obtained were annealed in high pressure oxygen atmosphere ($p$~60 atm) at 600 °C for 2 days.

In powder X-ray diffraction patterns taken with Fe*Kα* radiation, no impurity phases were detected for the samples sintered at 1150 °C and 1200 °C, while weak peaks from the $Ca_3Co_4O_9$ and other unknown impurity phase(s) were detected for the samples sintered at 1000 °C and 1100 °C. The oxygen numbers of the samples prepared at 1200 °C (and annealed in high oxygen pressure) were determined by the thermo gravimetric analysis (TGA) to be larger than 2.98 and here we use the oxygen number of 3. (Neutron structural analyses did not show significant oxygen deficiency, either.) To determine the Ca concentration $x$ of the $Pr_{1-x}Ca_xCoO_3$ phase in these samples, we applied the combined use of the X-ray and neutron Rietveld analyses as stated later. Powder neutron diffraction measurements were carried out using a high resolution powder diffractometer (HRPD) in JRR-3 of JAERI in Tokai at 10 K and room temperature for samples sintered at 1000 °C, 1100 °C, 1150 °C and 1200 °C with the nominal $x$ values of 0.5, and for a sample sintered at 1200 °C with the nominal $x$ value of 0.3. Rietveld analyses were carried out for these data by using the computer program Rietan 2000.[20] Although we could determine the oxygen position well in the refinements because of the relatively large scattering amplitude $b$ of oxygen atoms, the Ca concentration $x$ could not be determined precisely, because of the fact that $b$ of Pr nuclei ($b=4.4\times10^{-13}$cm) is nearly equal to that of Ca nuclei ($b=4.83\times10^{-13}$cm). To overcome this difficulty, we also took powder X-ray diffraction data and carried out the Rietveld analyses. In the analyses, we used the positional parameters determined by the neutron Rietveld analyses, and determined the Ca concentration $x$ for each sample. From the results, we found that the real $x$ values of the samples sintered at 1200 ºC were equal to the nominal values within the error bars of 0.02. The real $x$ values of the samples sintered at 1000 ºC, 1100 ºC and 1150 ºC, which have the nominal $x$ values of 0.5 are 0.22±0.02, 0.40±0.01 and 0.48±0.02, respectively, indicating that the Ca atoms do not completely go to the Pr sites at the sintering temperature $T_a$ lower than 1200 ºC. In Table I, the parameters determined by the neutron and X-ray Rietveld analyses are summarized. (All samples have the orthorhombic unit cell, whose volume is described by $\sim\sqrt{2}a_p \times 2a_p \times \sqrt{2}a_p$ (space group $Pnma$); $a_p$ being the lattice constant of the cubic perovskite cell.) Although these parameters were determined with fixed oxygen number at 3.0, no significant change has been found, even when the oxygen number is optimized.

In Fig.1, the unit cell volumes $V_u$ of $Pr_{1-x}Ca_xCoO_3$ sintered at various temperatures are plotted against the real $x$ values determined by the Rietveld analyses. The data points taken for all samples are on a single line, which indicates that real $x$ values are well estimated by the above analyses. Among all the samples of $Pr_{1-x}Ca_xCoO_3$ prepared here, the transition has been observed at ambient pressure in transport and magnetic measurements only for samples sintered at 1200 ºC and with the nominal $x$ value of 0.5.

Polycrystalline samples of $(Pr_{1-y}R'_y)_{1-x}Ca_xCoO_3$ (R'=La, Nd, Sm, Tb and Y; $x$=0.3, 0.4 and 0.5; 0≤$y$≤0.2) were synthesized from the mixtures of $Pr_6O_{11}$, $R'_2O_3$ (R'=La, Nd, Sm and Y), $R'_4O_7$ (R'=Tb), $CaCO_3$ and $CoC_2O_4 \cdot 2H_2O$ with proper molar ratios. All mixtures were sintered at 1200 ºC and then annealed as described above. In every sample preparation, we have prepared samples with $x$=0.5 and $y$=0 simultaneously as the standard one, because physical parameters of the samples may depend on each preparation process. No impurity phase has been detected in these samples. We have measured the resistivity ρ of the standard sample in every series of the samples and confirmed that all the standard samples exhibit the transition. In Fig.2, the unit cell volumes of $(Pr_{1-y}R'_y)_{1-x}Ca_xCoO_3$ at room temperature are plotted against $y$ for various R'. For R'=La, whose ionic radius is larger than that of $Pr^{3+}$, the volume increases with $y$, while For R'=Nd, Sm, Tb and Y, whose ionic radii are smaller than that of $Pr^{3+}$, the volume decreases with $y$.[21]

Electrical resistivities ρ were measured by the standard four-terminal method by using an ac-resistance bridge on heating from 4.2 to 300 K. In the resistivity measurements, the external pressure $p$ was generated up to 15 kbar (at room temperature) by using a clamp cell with the fluid pressure medium, the mixture of Daphne Oil (Idemitsu) 7243 and 7343. The pressure decreases by about 2 kbar when the temperature is lowered from room temperature to 150 K. Below 150 K, the pressure is nearly $T$-independent. Further details of the measurements under pressure can be found in ref. 22. The $p$ values at room temperature are used in presenting the $T$ dependence of the resistivities ρ. Magnetic susceptibilities χ were measured by using a Quantum Design SQUID magnetometer under the magnetic fields $H$ of 0.1 T in the temperature range of 5-700 K. In the magnetization measurements, the pressure $p$ was generated up to about 8 kbar (at 7 K) by using a clamp cell with Daphne Oil 7343. The $p$ value was estimated by the measurements of the superconducting transition temperature $T_c$ of lead ($T_c$=7.19 K, $dT_c/dp$=-0.384 K/GPa).[23] The thermal dilatation measurements were carried out using the strain gauge, where the electrical resistance of the gauge was converted into the linear thermal expansion. Parameters of the conversion formula were determined by measuring the Cu standard specimen.[24] Powder X-ray diffraction measurements were also carried out in SPring-8 (beam line: BL02B2, BL10XU). The $^{59}$Co-NMR measurements were carried out by a *p*/2-*t*-*p* phase-cycled pulse sequence with fixed *t*. The spectral intensity was estimated by integrating the whole spectral weights. The $T_2$ correction was made.

## 3. Experimental Results and Discussion

### 3.1 T-x phase diagram and the electronic states of $Pr_{1-x}Ca_xCoO_3$

Figure 3 shows the electrical resistivities ρ of $Pr_{1-x}Ca_xCoO_3$ with $x$=0.2 and 0.5 sintered at 1200 ºC taken under various values of the applied pressure $p$. At ambient pressure, the transition occurs in the sample with $x$=0.5, but does not occur in the sample with $x$=0.2. Under high pressure, the transition is observed in both samples. The transition temperature $T_s$, which is defined as the temperature of the inflection point of the logρ-$T$ curve, increases with increasing $p$.

Figure 4 shows $T_s$-$x$ phase diagram of $Pr_{1-x}Ca_xCoO_3$ for various values of $p$. Data obtained for the samples sintered at 1200 ºC and 1000 ºC are plotted as functions of the real $x$ value by solid and open circles, respectively, where for the real $x$ values of the 1000 ºC sintered samples, we used the values determined by the Rietveld refinements. The pressure dependence of the $T_s$-$x$ curve indicates that the curves change smoothly from the one at ambient pressure to the one at $p$=15 kbar as $p$ increases: The region of the low temperature phase in the phase diagram becomes wider as $p$ becomes larger and we can consider that there exist only one mechanism of the transition in all the region of $p$.

Because the transition occurs in the wide region of $x$ under high pressure, the mechanism does not seem to be an order-disorder type one, which is consistent with the fact that the low temperature structure deduced at ambient pressure for a sample sintered at 1200 ºC with $x$=0.5 by the neutron Rietveld analysis[15] does not suggest a charge-ordering below $T_s$ (See Table I, too.).

In Fig. 5, the magnetizations $M$ divided by the magnetic field $H$ measured under high pressure up to about 8 kbar for the sample of $Pr_{0.6}Ca_{0.4}CoO_3$ sintered at 1200 ºC are plotted against $T$, where data taken with the condition of the zero field cooling are shown. (Inset shows the $M$-$T$ curves with the larger vertical scale. In this case, both data taken with the conditions of the zero field cooling and the field cooling are shown.) The $p$ values in the figure are estimated at 7 K. The abrupt increase of $M/H$ at $T_s$ with increasing $T$ under high pressure can be seen, as was reported at ambient pressure by Tsubouchi *et al.* for their sample of $Pr_{0.5}Ca_{0.5}CoO_3$ sintered at 1200 ºC. Ferromagnetic moment which appears at low temperatures in the low pressure region vanishes with increasing $p$ and $T_s$ increases. These results can be understood by considering that at low temperatures, the electrons of $Co^{3+}$ ions in $Pr_{0..6}Ca_{0.4}CoO_3$ go into the LS state with increasing $p$.

*3.2 Transport studies on $(Pr_{1-y}R'_y)_{1-x}A_xCoO_3$ {R'= (rare earth elements and Y); A=( Ca, Ba and Sr) }*

We have extended the studies to the system $(Pr_{1-y}R'_y)_{1-x}Ca_xCoO_3$ (R'=rare earth elements and Y) at ambient pressure. In Fig. 6(a), ρ of $(Pr_{1-y}Y_y)_{0.5}Ca_{0.5}CoO_3$ (0≤$y$≤0.2) at ambient pressure are shown against $T$. $T_s$ increases with increasing $y$ as in the case of the pressure effect. Figure 6(b) shows the $T$ dependence of ρ of $(Pr_{1-y}Tb_y)_{0.7}Ca_{0.3}CoO_3$ (0≤$y$≤0.2) at ambient pressure. The sample with $y$=0 does not exhibit the transition, but we find that the transition is induced by the small amount (10 %) of the Pr substitution with Tb.

In Fig. 7, the transition temperatures $T_s$ of $(Pr_{1-y}R'_y)_{1-x}Ca_xCoO_3$ ($x$=0.5 for R'=La, $La_{0.2}Nd_{0.8}$, Nd, Sm, Tb and Y, and $x$=0.3 for R'=Tb and Y) defined as the inflection points of the logρ-$T$ curves are shown against $y$ at ambient pressure. Generally speaking, the substitution of Pr with elements whose ionic radii are smaller than $Pr^{3+}$ enhances $T_s$ and the substitution with larger ions suppresses $T_s$. An exception is found for R'=Nd, in which $T_s$ slightly decreases with increasing $y$, even though the ionic radius of $Nd^{3+}$ is smaller than that of $Pr^{3+}$. It is probably due to an effect of the randomness introduced by the substitution. The substitution with $La_{0.2}Nd_{0.8}$, whose average ionic radius is equal to that of $Pr^{3+}$ also suppresses $T_s$. We have found that just the 5 % substitution of Ca with Sr suppresses the transition completely. The $T_s$ values determined for the systems derived by the (Pr→R') and (Ca→Sr) substitutions and those determined for $Pr_{1-x}Ca_xCoO_3$ under the external pressure indicate that $T_s$ is sensitive to the unit cell volume (and possibly to the volume of the $CoO_6$ octahedra). This sensitive volume dependence of $T_s$ can be understood by considering that the decrease of the volume of the $CoO_6$ octahedra induces the increase of the crystal field splitting energy $\Delta_c$. However, we may have to consider, as is discussed later, effects of the possible change of the Co-O-Co transfer energy, which, we think, affect the $T_s$ value, too.

*3.3 Role of Pr in the occurrence of the transition*

As was reported in our previous paper, the transition seems to be observed only in Pr-Ca system.[8] In the present work, we have searched the transition again in the samples of $R_{1-x}Ca_xCoO_3$ with (R, $x$)=(La, 0.5), (Nd, 0.3) and ($La_{0.2}Nd_{0.8}$, 0.5) sintered at 1200 ºC. The ratio of the La and Nd atoms in the last set was chosen to adjust their average ionic radius to that of Pr. We have also searched the transition for the samples of $Pr_{1-x}A_xCoO_3$ with (A, $x$)=(Ba, 0.5), (Sr, 0.1), (Sr, 0.2) and (Sr, 0.5) sintered at 1200 ºC, but no indication of the resistivity anomaly under high pressure up to 15 kbar has been found. We speculate that the transition in $Pr_{1-x}Ca_xCoO_3$ might occur by the mechanism of the hole trapping in the Pr4f-O2p$_\pi$ hybridized orbital, which was first proposed by Fehrenbacher-Rice to explain the suppression of the superconducting transition temperature $T_c$ by Pr atoms in $R_{1-x}Pr_xBa_2Cu_3O_{6+y}$ (R=rare earth elements and Y). [25]

*3.4 Discussion of the electronic or spin state of Co from the structural aspect*

The electronic state of the present Co oxides is rather closely related to their local structures. Typical examples of the relationship between the spin state of $Co^{3+}$ ions and local structures surrounding them can be found in our works on $RBaCo_2O_{5+\delta}$ carried out by means of the single crystal neutron diffraction.[26,27] In the works, the spin states of the Co ions in $CoO_6$ and $CoO_5$ polyhedra have been determined. To argue the relationship between the spin state of Co ions and local structures surrounding them in more detail, we plot in Fig. 8, as functions of the real Ca concentration $x$, the unit cell volume $V_u$, the volume of $CoO_6$ octahedra, $V_o$ and the Co-O-Co bond angle $\alpha$ of $Pr_{1-x}Ca_xCoO_3$ determined at ambient pressure by the neutron and X-ray Rietveld analyses stated above at 10 K and at room temperature. (At ambient pressure the transition appears only in the sample with $x$=0.5. The data at room temperature for $x$=0 are from ref. 12.) Both $V_u$ and $V_o$ decrease significantly with increasing $x$ in the paramagnetic region of the relatively large $x$. From the $x$ dependence of these volumes, we expect there that the crystal field splitting $\Delta_c$ increases with increasing $x$. However, $^{59}$Co-NMR results indicate that $\delta E$ decreases with increasing $x$. Then, before the discussion on the $x$ dependence of $\delta E$, let us describe the NMR results.

In the inset of Fig. 9, the $^{59}$Co-NMR spectra taken for $x=0$ are shown. With increasing $x$, the peak intensity decreases significantly and the spectra are broadened within the $H$ region between 1.5~1.9 T. The integrated intensity $I$ was estimated by integrating the spectral weight over this region. Figure 9 shows the $T$ dependence of $I \times T$ obtained for several samples of $Pr_{1-x}Ca_xCoO_3$ sintered at 1200 °C. (Only the exception is the one with $x=0.48$. It is sintered at 1150 °C). The quantity $I \times T$ is considered to be proportional to the number of $Co^{3+}$ ions in the LS state, which can be rationalized by the following fact. $I \times T$ decreases with increasing $x$ (except for the sample with $x=0.5$ at $T<T_s$), and the $T$ dependence can be fitted, as shown in the figure by the solid line for each $x$, by the number of the LS state of $Co^{3+}$ ions calculated for the level scheme of the LS, IS and HS states obtained by the analyses of the $T$ dependence of the magnetic susceptibility $\chi$ of the corresponding sample. (See Fig. 10 and Table II. The deviations of the observed data from the broken lines found in Fig. 9 are due to the partial occurrence of the ferromagnetism, which disturbs the observation of the NMR signal of $Co^{3+}$ in the neighborhood region of the ferromagnetic parts.) The abrupt increase of $I \times T$ at $T_s$ observed for $x=0.5$ with decreasing $T$ is consistent with the abrupt increase of the numbers of $Co^{3+}$ ions in the LS state.

The $T$ dependence of the inverse magnetic susceptibilities $\chi^{-1}$ of the samples with $x=0.0$, 0.3 and 0.48, which do not exhibit the transition, are shown in Fig. 10 together with calculated ones (solid curves). In the calculations, we have approximated the value of the $Pr^{3+}$ magnetic moment to be 3.41 $\mu_B$.[28] Although this moment value is not a good approximation in the low $T$ region (<100 K), because the ions tend to approach the nonmagnetic state, we can reasonably describe the magnetic susceptibilities of $Pr^{3+}$ ions above 100 K, and the energies $\delta E$ of the IS and HS states of $Co^{3+}(Co^{4+})$ ions with respect to the LS ground state level of $Co^{3+}(Co^{4+})$ ions can be determined by fitting the calculated $\chi^{-1}$-$T$ curve to the observed one for each sample. (In the fittings, the degeneracy of each level is considered. The HS state of $Co^{4+}$ was ignored, because the energy is rather high and its degeneracy is smaller than other excited states.) The results are shown in Table II. Although the fittings may not be precise, they present, we think, a qualitative interpretation of the data. In particular, they give a reasonable idea that the NMR spectra are from the $Co^{3+}$ ions. They also give us the rough values of the energy level schemes of the spin states of $Co^{3+}$ and $Co^{4+}$.

The $x$ dependence of $V_u$ and $V_o$ and the NMR results suggest that $\delta E$ depends on not only $\Delta_c$ (and of course Hund coupling energy $J_H$) but also certain other parameters. We presume that one of possible candidates of such parameters is, related to a kind of

"kinetic energy", say $K$, which depends on the transfer integral $t$ and the hole concentration $n$

In Fig. 11(a), the unit cell volumes $V_u$ of $Pr_{1-x}Ca_xCoO_3$ are shown for $x$=0.3 and 0.5 by open and closed circles, respectively, determined by the neutron diffraction studies, where the $T$ dependence of the volumes estimated from the thermal dilatation data taken at ambient pressure for polycrystalline samples of $(Pr_{1-y}Tb_y)_{1-x}Ca_xCoO_3$ with several sets of ($x, y$) are also shown. We can find rather good agreements of both kinds of data, which indicates that $V_u$ changes abruptly at the transition temperatures $T_s$. Figures 11(b) and 11(c) show the $T$ dependence of $V_o$ and $\alpha$, respectively. From the figures, it can be found that the transition is primarily due to the abrupt decrease of $\alpha$ or the abrupt increase of the tilting angle of the $CoO_6$ octahedra at $T_s$, which is accompanied by the release of the volume contraction of the octahedra.

We realize that in Fig. 8 the differences of the unit cell volumes $V_u$ and the bond angles $\alpha$ between the two temperatures, room temperature and 10 K for the sample ($x$=0.5), in which the transition can be seen, are significantly larger than those of the other samples, while the difference of the volumes of the $CoO_6$ octahedra, $V_o$ between the two temperatures for $x$=0.5 is much smaller than those of the other samples. The result can be easily understood from the $T$ dependence of $V_u$, $V_o$ and $\alpha$.

We have tried to determine the critical value $V_{uc}$ of the unit cell volume $V_u$ at which the transition takes place with decreasing $T$ for $(Pr_{1-y}Tb_y)_{1-x}Ca_xCoO_3$ in the following way: As can be seen in Fig. 11(a), the changing rate of $V_u$, $dV_u/dT$, is almost independent of $x$ above $T_s$. Figure 12(a) shows the $T$ dependence of $V_u$ determined for a 1150 °C sintered sample with the nominal $x$ value of 0.5 (real $x$ value of ~0.48) by the X-ray diffraction measurements under the pressure of 12.5 kbar at SPring-8 (BL10XU). Under high pressure, $dV_u/dT$ is larger than that at ambient pressure. Figure 12(b) show the $p$ dependence of $V_u$ estimated at room temperature for the sample of $Pr_{0.7}Ca_{0.3}CoO_3$ sintered at 1150 °C by taking the X-ray diffraction peaks under pressure at SPring-8. In the studies, we have found by the detailed structure analyses at room temperature up to 20 kbar that as shown in Fig. 13, $V_o$ and $V_u$ decrease with increasing $p$ with the bond angle $\alpha$ being kept nearly constant.

Here, to estimate $V_{uc}$ values we use following assumptions. (1) $dV_u/dT$ above $T_s$ is independent of $x$, (2) it increases linearly in $p$ and (3) the value of $dV_u/dp$ at room temperature is $x$-independent. (This does not change the results so significantly.) Then, $V_{uc}$ determined for each sample are summarized in Fig. 14 in the form of $T_s$-$V_{uc}$ for $(Pr_{1-y}Tb_y)_{1-x}Ca_xCoO_3$ samples with varying $y$ at several fixed values of $x$ at ambient pressure and for $Pr_{1-x}Ca_xCoO_3$ samples with varying $p$ at two fixed values of $x$.

What information can we extract from these results? First, we consider a sample series with fixed $x$ and varying $y$, in which all samples have same electron numbers.

With increasing $y$, $V_u$ and $V_o$ above $T_s$ decreases and $\Delta_c$ increases. The Co-O-Co bond angle $\alpha$ may also decrease, which causes the decrease of the transfer energy $t$. All of these changes enhance $T_s$ (curves b-d in Fig. 13). At $T_s$, the abrupt decrease of $\alpha$ or the abrupt increase of the tilting angle of the octahedra takes place and the volume contraction of the octahedral is released (See Fig. 11 (b)). By this reduction of $\alpha$, the transfer energy between Co atoms in the $t_{2g}$ orbital is suddenly reduced. In this sense, we can say that to stabilize the LS state, it seems to be more important to reduce the $t$ value than to increase $\Delta_c$.

For a fixed $y$, the $T_s$-increase with increasing $x$ or with decreasing $V_u$, is much slow (curve a) as compared with the case of the curves b-d. Phenomenologically, the carrier doping suppresses the transition temperature $T_s$ (as can be directly found by comparing the curves b-d). Based on this fact, we think that the increase of the "kinetic energy" $K$ suppresses the $T_s$ increase, which is consistent with the idea stated just above that the reduction of $\alpha$ causes the increase of $T_s$

The $p$ dependence of $V_{uc}$ plotted in Fig. 14 for $(x, y)=(0.3, 0)$ and $(0.5, 0)$ may not be so clearly understood by the above idea. The slope $|dT_s/dV_u|$ is small in the low $p$ region ($p<10$ kbar) and it becomes larger as $p$ exceeds 10 kbar. If we apply the above idea that $V_u$, $V_o$ and the "kinetic energy" determine the transition temperature $T_s$, the small value of $|dT_s/dV_u|$ at $p<10$ kbar should be considered to be caused by the increase of the "kinetic energy" with increasing $p$. On this point, we note that the bonding angle $\alpha$ does not exhibit a significant change with $p$: The tolerance factor defined as $(r_R+r_O)/\sqrt{2}(r_{Co}+r_O)$ by using the average ionic radii $r_M$ (M=R, O and Co) of corresponding sites does not sensitively depend on $p$, because the contraction with $p$ occurs in both the R-O and $CoO_2$ planes. Then, the tilting of the $CoO_6$ octahedra which takes place to adjust the difference between the lengths $(r_R+r_O)/\sqrt{2}$ and $(r_{Co}+r_O)$ does not sensitively change its magnitude with $p$, indicating that by the external pressure the volume contraction of the $CoO_6$ octahedra is induced with $\alpha$ being kept nearly constant. It has been experimentally confirmed as stated above. (This may be a notable difference of the pressure effect from effects of the substitution of Pr.) Then, the transfer energy $t$ of the electrons in the $t_{2g}$ orbitals increases, which may explain the why $|dT_s/dV_u|$ is small at $p<10$ kbar as is found in Fig.14. (One might think that the $p$ dependence of the resistivities shown in Fig. 3 contradicts the idea, because the resistivities above $T_s$ do not decrease with $p$. However, in this case, we have to consider the contribution of the electrons in the $e_g$ orbitals to the conductivity: The conductivity decreases with $p$ due to the reduction of the electron number caused by the increase of $\Delta_c$.)

We have discussed on what parameters are important to control the electronic state of Co-oxides, on the basis of the results on the samples of $(Pr_{1-y}R'_y)_{1-x}A_xCoO_3$

{R'= (rare earth elements and Y); A=(Ca, Ba and Sr) }. The relationship between the $V_u$, $V_o$ and Co-O-Co bond angle (or tilting angle of the octahedra) and the stability of the LS ground state has been discussed, though it is still not easy to have a quantitative understandings what factors are important to control the electronic state of the system.

## 4. Conclusion

Various kinds of studies have been carried out mainly on the phase transition found in $(Pr_{1-y}R'_y)_{1-x}A_xCoO_3$ {R'= (rare earth elements and Y); A=(Ca, Ba and Sr) }. The $T$-$x$ phase diagram has been constructed at $y=0$ for various values of $p$ and the transition has been found not to be an order-disorder type one. For the occurrence of the transition in $(Pr_{1-y}R'_y)_{1-x}A_xCoO_3$, both the Pr and Ca atoms seem to be necessary. The intimate relationship between the local structure around the Co ions and the electronic (or spin) state of $Co^{3+}$ ions has been discussed: The role of the carriers introduced by the doping of the A atoms, is also discussed. Results of the analyses of the $^{59}$Co-NMR spectra and the magnetic susceptibilities $\chi$ indicate that the energy separations among the levels of the different spin states of $Co^{3+}$ and $Co^{4+}$ are roughly estimated. All these results present useful information on what parameters are essential to control the electronic state.

Acknowledgements – Work at the JRR-3 was performed within the frame of JAERI Collaborative Research Program on Neutron Scattering. The synchrotron radiation experiments were performed at the BL10XU in the SPring –8 with the approval of the Japan Synchrotron Radiation Research Institute. (Proposal No. 2003A-5004-LD-np). The work is supported by Grants-in-Aid for Scientific Research from the Japan Society for the Promotion of Science (JSPS) and by Grants-in-Aid on priority area from the Ministry of Education, Culture, Sports, Science and Technology.

Figure captions

Fig. 1  Unit cell volumes $V_u$ of $Pr_{1-x}Ca_xCoO_3$ sintered at various temperatures are plotted against the real $x$ value. The circles and the squares show the unit cell volumes for the samples sintered at $T_a=1200$ ºC and $T_a<1200$ ºC, respectively. The filled and open symbols show the unit cell volumes deduced by the powder X-ray and neutron Rietveld refinements, respectively.

Fig. 2  Unit cell volumes $V_u$ of the 1200 °C sintered samples of $(Pr_{1-y}R'_y)_{1-x}Ca_xCoO_3$ with $x=0.3$ and 0.5 are plotted against $y$ for various R' atoms.

Fig. 3  Electrical resistivities ρ of the 1200 °C sintered samples of $Pr_{1-x}Ca_xCoO_3$ with $x=0.2$ and 0.5 taken under various values of the applied pressure are shown against $T$.

Fig. 4  $T_s$-$x$ phase diagram of $Pr_{1-x}Ca_xCoO_3$ for various values of $p$. The closed and open circles are the data for the samples sintered at 1200 ºC and 1000 ºC, respectively, where the real $x$ values are used.

Fig. 5  Main panel indicates the $T$-dependence of the magnetic susceptibilities $c$ of $Pr_{0.6}Ca_{0.4}CoO_3$ measured with the magnetic field $H=0.1$ T under various values of the applied pressure. Inset shows the magnetization $M$ of $Pr_{0.6}Ca_{0.4}CoO_3$ plotted against $T$ in the lager vertical scale. The data taken under the zero-field-cooling condition have smaller values in the low temperature region than those taken under the field-cooling condition.

Fig. 6(a)  Electrical resistivities ρ of $(Pr_{1-y}Y_y)_{0.5}Ca_{0.5}CoO_3$ ($0 \leq y \leq 0.2$) at ambient pressure are shown against $T$.

Fig. 6(b)  Electrical resistivities ρ of $(Pr_{1-y}Tb_y)_{0.7}Ca_{0.3}CoO_3$ ($0 \leq y \leq 0.2$) at ambient pressure are shown against $T$.

Fig. 7  Transition temperatures $T_s$ of the $(Pr_{1-y}R'_y)_{1-x}Ca_xCoO_3$ samples with $x=0.5$ at ambient pressure are shown for various R' atom species. They were sintered at 1200 ºC. Inset shows similar data for $x=0.3$.

Fig. 8  Unit cell volumes $V_u$ (top), volumes of $CoO_6$ octahedra, $V_o$ (middle) and Co-O-Co bond angles α (bottom) of $Pr_{1-x}Ca_xCoO_3$ at 10K (open symbols) and room temperature (closed ones) are shown against $x$. They were determined by powder neutron Rietveld refinements at ambient pressure (space group *Pnma*; data for $x=0$ at room temperature are from H.W.Brinks *et al.*[12] No data are available at 10 K for $x=0$.) In the bottom figure, two crystallographycally distinct sites (n=1 and 2) are shown.

Fig. 9  Integrated intensities of $^{59}$Co-NMR spectra of $Pr_{1-x}Ca_xCoO_3$ multiplied by $T$ are shown for various $x$ values. The solid and broken lines show the $T$ dependences calculated by using the energy separations of the spin state levels

of $Co^{3+}$ and $Co^{4+}$ ions obtained by the fitting to the susceptibility data (see Fig. 10), which are shown in Table II. The deviations of the data from the broken lines are due to the partial occurrence of the ferromagnetism, which disturbs the observation of the NMR signal of Co-nuclei in the neighborhood of the ferromagnetic region. The long dashed line is just guide for the eye for $x$=0.5. Inset shows an example of the $^{59}$Co-NMR spectra, which was observed for $x$=0 at 100 K.

Fig. 10 Magnetic susceptibilities $\chi$ of $Pr_{1-x}Ca_xCoO_3$ are plotted against $T$ for three $x$ values. The solid lines are calculated results for the parameters of $\delta E$ values shown in Table II.

Fig. 11(a) $T$-dependence of the unit cell volume $V_u$ of several samples of $(Pr_{1-y}Tb_y)_{1-x}Ca_xCoO_3$ at ambient pressure obtained by the measurements of the linear thermal dilatation are shown by the solid and broken curves for various sets of ($x$, $y$). Open and solid circles show the unit cell volumes determined by the neutron diffraction for the samples with ($x$, $y$)=(0.3, 0) and (0.5, 0) at several temperatures. (b) $T$ dependence of $V_o$ is shown for ($x$, $y$)=(0.3, 0) and (0.5, 0). The dotted line is just for the eye. (c) $T$ dependence of $\alpha$ is shown for ($x$, $y$)=(0.3, 0) and (0.5, 0). The dotted line is just for the eye. The crystallographically distinct oxygen sites are indicated by n=1 and 2.

Fig.12 (a) $T$-dependence of the unit cell volume determined by the powder X-ray diffraction at SPring-8 (BL10XU) at $p \approx 12.5$ kbar for $Pr_{0.52}Ca_{0.48}CoO_3$ sintered at 1150 $^o$C is shown. (b) The $p$ dependence of the unit cell volume determined by powder X-ray diffraction in SPring-8 (BL10XU) at room temperature is shown for $Pr_{0.7}Ca_{0.3}CoO_3$. They were sintered at 1150 $^o$C.

Fig.13 Pressure dependences of $V_u$, $V_o$ and $\alpha$ obtained by the X-ray Rietveld refinements at room temperature are shown for $Pr_{0.7}Ca_{0.3}CoO_3$.

Fig.14 $V_{uc}$-$T_s$ phase diagram of $(Pr_{1-y}Tb_y)_{1-x}Ca_xCoO_3$: The data are taken for various series with fixed values of $x$=0.3, 0.4 and 0.5 and varying $y$ at ambient pressure(curves b-d), and those taken with fixed values of $x$=0.3 and 0.5 and $y$=0 and with varying applied pressure $p$(curves e and f) are shown. The curve a indicates the data taken with $y$=0.1 and with varying $x$. ($V_{uc}$ values are defined as the unit cell volume at $T_s$+0.)

Table I. Structural parameters of the samples of $Pr_{1-x}Ca_xCoO_3$ sintered at temperatures $T_a$ = 1000 °C, 1100 °C, 1150 °C and 1200 °C determined by the combined use of neutron and X-ray Rietveld analyses are shown (Space group $Pnma$). Both the nominal and real $x$ values are listed. See text for details.

| $T_a$(°C) | 1000 | 1000 | 1100 | 1100 | 1150 | 1150 | 1200 | 1200 | 1200 | 1200 | 1200 | 1200 | 1200 |
|---|---|---|---|---|---|---|---|---|---|---|---|---|---|
| nominal $x$ | 0.5 | 0.5 | 0.5 | 0.5 | 0.5 | 0.5 | 0.5 | 0.5 | 0.5 | 0.5 | 0.5 | 0.3 | 0.3 |
| real $x$ | 0.22±0.02 | 0.22±0.02 | 0.40±0.01 | 0.40±0.01 | 0.48±0.02 | 0.48±0.02 | 0.50±0.02 | 0.50±0.02 | 0.50±0.02 | 0.50±0.02 | 0.50±0.02 | 0.30±0.02 | 0.30±0.02 |
| $T$(K) | 295 | 10 | 295 | 10 | 295 | 10 | 295 | 100 | 60 | 10 | 10 | 295 | 10 |
| $a$(Å) | 5.3540(5) | 5.3360(6) | 5.3521(2) | 5.3389(2) | 5.3445(1) | 5.3331(3) | 5.3431(1) | 5.3346(1) | 5.3270(1) | 5.3254(2) | — | 5.3526(1) | 5.3369(1) |
| $b$(Å) | 7.5783(8) | 7.5475(8) | 7.5678(3) | 7.5378(2) | 7.5545(2) | 7.5219(4) | 7.5470(1) | 7.5141(1) | 7.4696(2) | 7.4675(2) | — | 7.5714(2) | 7.5394(1) |
| $c$(Å) | 5.3765(6) | 5.35822(6) | 5.3629(2) | 5.3440(2) | 5.3514(2) | 5.3310(4) | 5.3407(1) | 5.3164(1) | 5.2698(1) | 5.2681(2) | — | 5.3650(1) | 5.3461(1) |
| $V$(Å$^3$) | 218.147(13) | 215.793(14) | 217.217(5) | 215.062(4) | 216.063(4) | 213.853(8) | 215.360(2) | 213.106(2) | 209.688(4) | 209.499(4) | — | 217.426(3) | 215.111(2) |
| $x$ [Pr,Ca;4c] | 0.5296(10) | 0.5319(10) | 0.5287(5) | 0.5331(5) | 0.5297(4) | 0.5344(5) | 0.5320(3) | 0.5367(3) | 0.5449(3) | 0.5455(3) | — | 0.5299(3) | 0.5325(3) |
| $z$ [Pr,Ca;4c] | 0.4949(12) | 0.4943(12) | 0.4971(8) | 0.4949(6) | 0.4960(6) | 0.4957(7) | 0.4928(5) | 0.4921(5) | 0.4912(4) | 0.4910(4) | — | 0.4948(4) | 0.4947(3) |
| $B$ [Pr,Ca;4c] | 0.65(10) | 0.35(7) | 0.70(4) | 0.35(3) | 0.74(4) | 0.39(4) | 0.66(3) | 0.49(3) | 0.44(3) | 0.26(2) | — | 0.64(2) | 0.35(2) |
| $B$ [Co;4b] | 0.32(11) | 0.03(9) | 0.32(5) | 0.14(5) | 0.36(5) | 0.15(5) | 0.36(4) | 0.31(4) | 0.18(5) | 0.03(4) | — | 0.28(3) | 0.03(2) |
| $x$ [O(1);4c] | -0.0074(16) | -0.0072(16) | -0.0109(4) | -0.0102(4) | -0.0105(3) | -0.0102(5) | -0.0082(3) | -0.0097(2) | -0.0140(2) | -0.0139(2) | — | -0.0079(2) | -0.0082(2) |
| $z$ [O(1);4c] | 0.5661(18) | 0.5673(16) | 0.5660(5) | 0.5680(4) | 0.5655(4) | 0.5680(6) | 0.5690(4) | 0.5698(3) | 0.5788(3) | 0.5809(3) | — | 0.5655(3) | 0.5676(3) |
| $B$ [O(1);4c] | 0.60(11) | 0.35(9) | 0.72(5) | 0.38(4) | 0.72(4) | 0.37(3) | 0.64(3) | 0.52(3) | 0.48(3) | 0.18(2) | — | 0.48(3) | 0.27(2) |
| $x$ [O(2);8d] | 0.2843(10) | 0.2846(10) | 0.2858(3) | 0.2859(3) | 0.2860(2) | 0.2871(3) | 0.2850(2) | 0.2874(2) | 0.2928(2) | 0.2933(2) | — | 0.2849(2) | 0.2852(1) |
| $y$ [O(2);8d] | 0.0351(8) | 0.0361(8) | 0.0349(2) | 0.0358(2) | 0.0353(2) | 0.0368(2) | 0.0339(1) | 0.0366(1) | 0.0408(1) | 0.0407(1) | — | 0.0349(1) | 0.0362(1) |
| $z$ [O(2);8d] | 0.2836(8) | 0.2849(12) | 0.2839(3) | 0.2850(3) | 0.2850(3) | 0.2878(3) | 0.2864(2) | 0.2890(2) | 0.2947(2) | 0.2945(2) | — | 0.2842(2) | 0.2854(2) |
| $B$ [O(2);8d] | 0.60(8) | 0.35(8) | 0.76(3) | 0.39(3) | 0.78(3) | 0.42(3) | 0.57(2) | 0.53(2) | 0.43(2) | 0.15(2) | — | 0.64(2) | 0.32(2) |
| $R_{wp}$ (%) | 8.07 | 8.35 | 5.93 | 6.20 | 5.50 | 6.52 | 5.36 | 5.62 | 6.27 | 5.93 | — | 4.51 | 5.00 |
| $R_e$ (%) | 6.16 | 5.28 | 4.60 | 4.64 | 4.60 | 4.65 | 4.98 | 5.34 | 5.38 | 5.06 | — | 4.44 | 4.73 |
| $R_I$ (%) | 4.98 | 4.83 | 3.25 | 3.66 | 3.47 | 3.37 | 2.09 | 2.03 | 1.63 | 2.57 | — | 1.77 | 1.74 |
| $R_f$ (%) | 3.08 | 2.73 | 1.93 | 2.34 | 2.04 | 1.87 | 1.83 | 1.39 | 1.13 | 1.25 | — | 1.16 | 1.14 |

Table II. Energies $\delta E$ of the excited states of $Co^{3+}(Co^{4+})$ ions in $Pr_{1-x}Ca_xCoO_3$ ($x$=0.0, 0.3 and 0.48) with respect to the LS ground state level of $Co^{3+}(Co^{4+})$ ions are shown. They were estimated by fitting the calculated $\chi^{-1}$-$T$ curves to the observed ones. See text for details. (For $x$=0.0, the value of the IS state of $Co^{3+}$ was estimated from the $T$ dependence of the NMR longitudinal relaxation rate $1/T_1$)

| $x$ | $\delta E$ (K) | | |
|---|---|---|---|
| | $Co^{3+}$:IS | $Co^{3+}$:HS | $Co^{4+}$:IS |
| 0.0 | 1100 (±100) | 1600 (±100) | - |
| 0.3 | 300 (±20) | 750 (±50) | 300 (±50) |
| 0.48 | 180 (±20) | 600 (±50) | 200 (±50) |

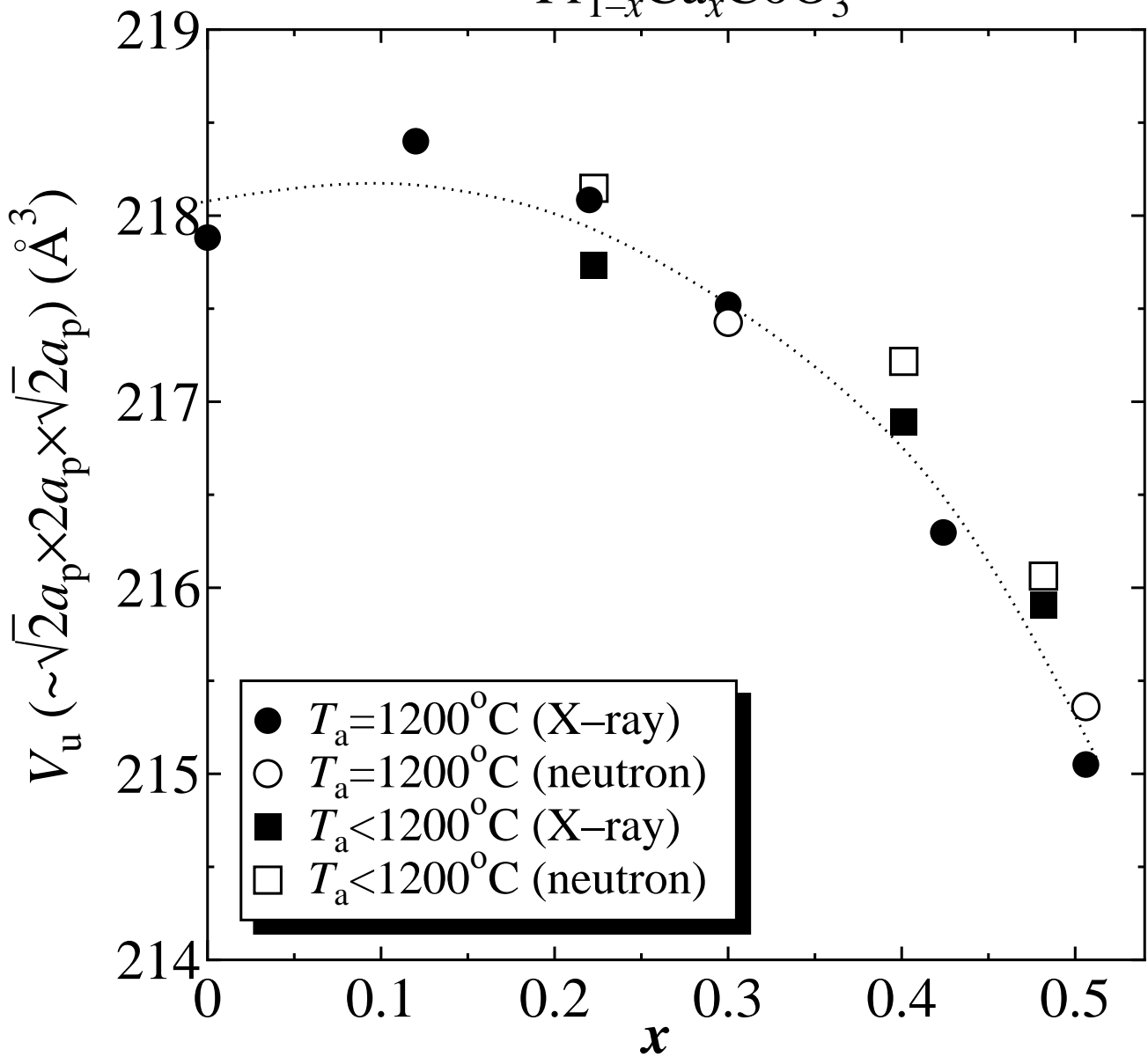

Fig.1
Fujita et al.

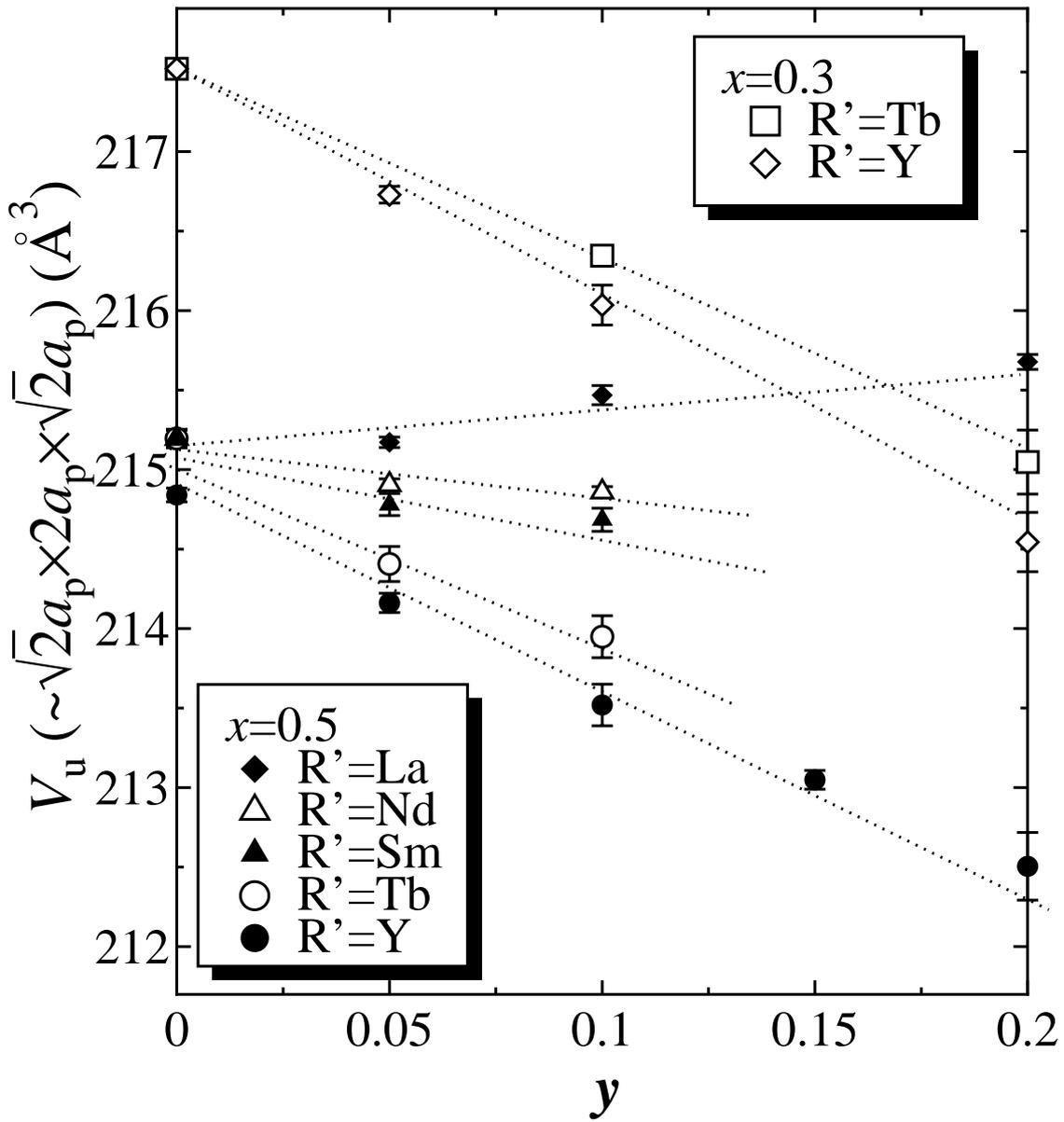

Fig.2
Fujita et al.

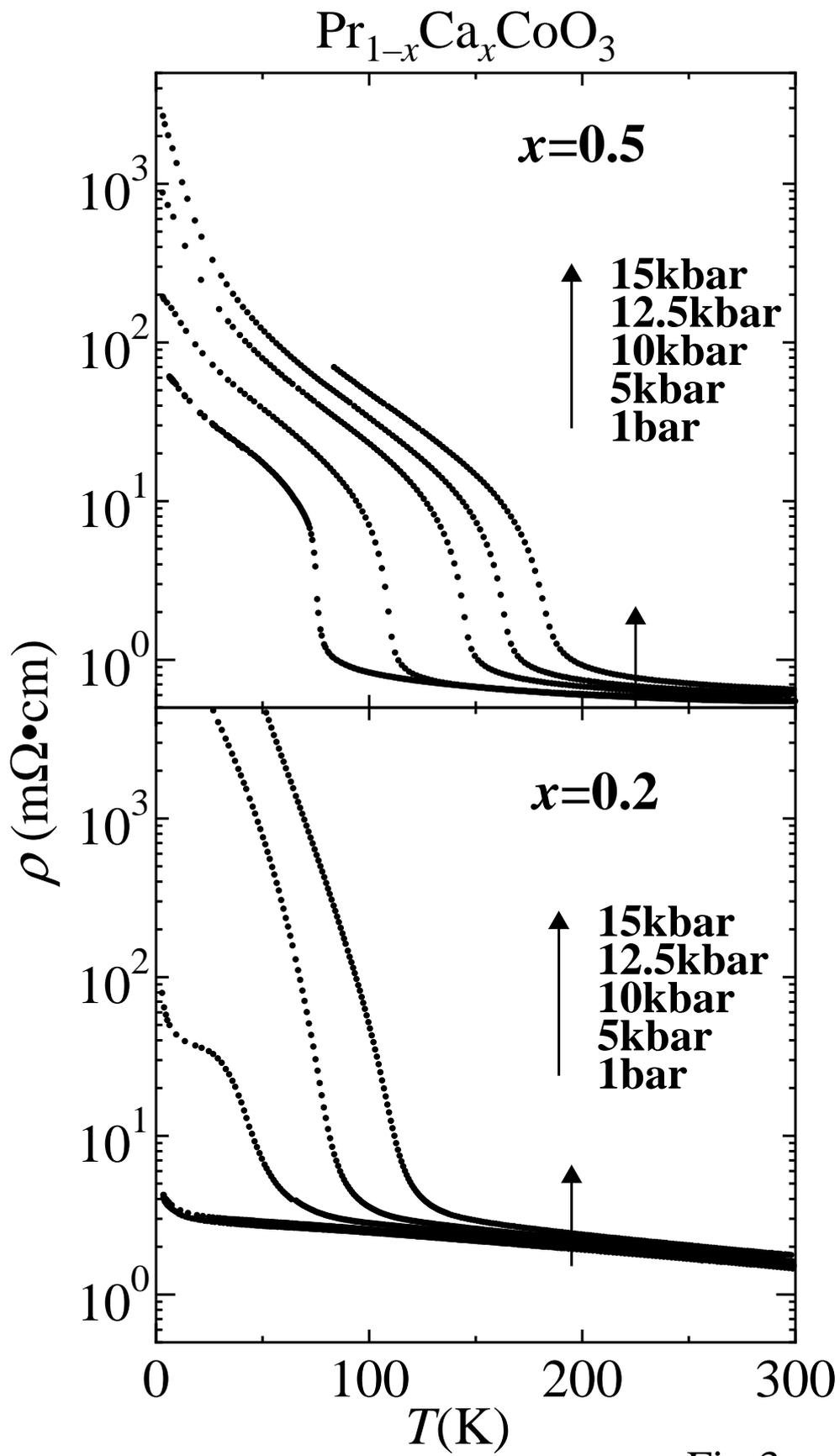

Fig.3
Fujita *et al.*

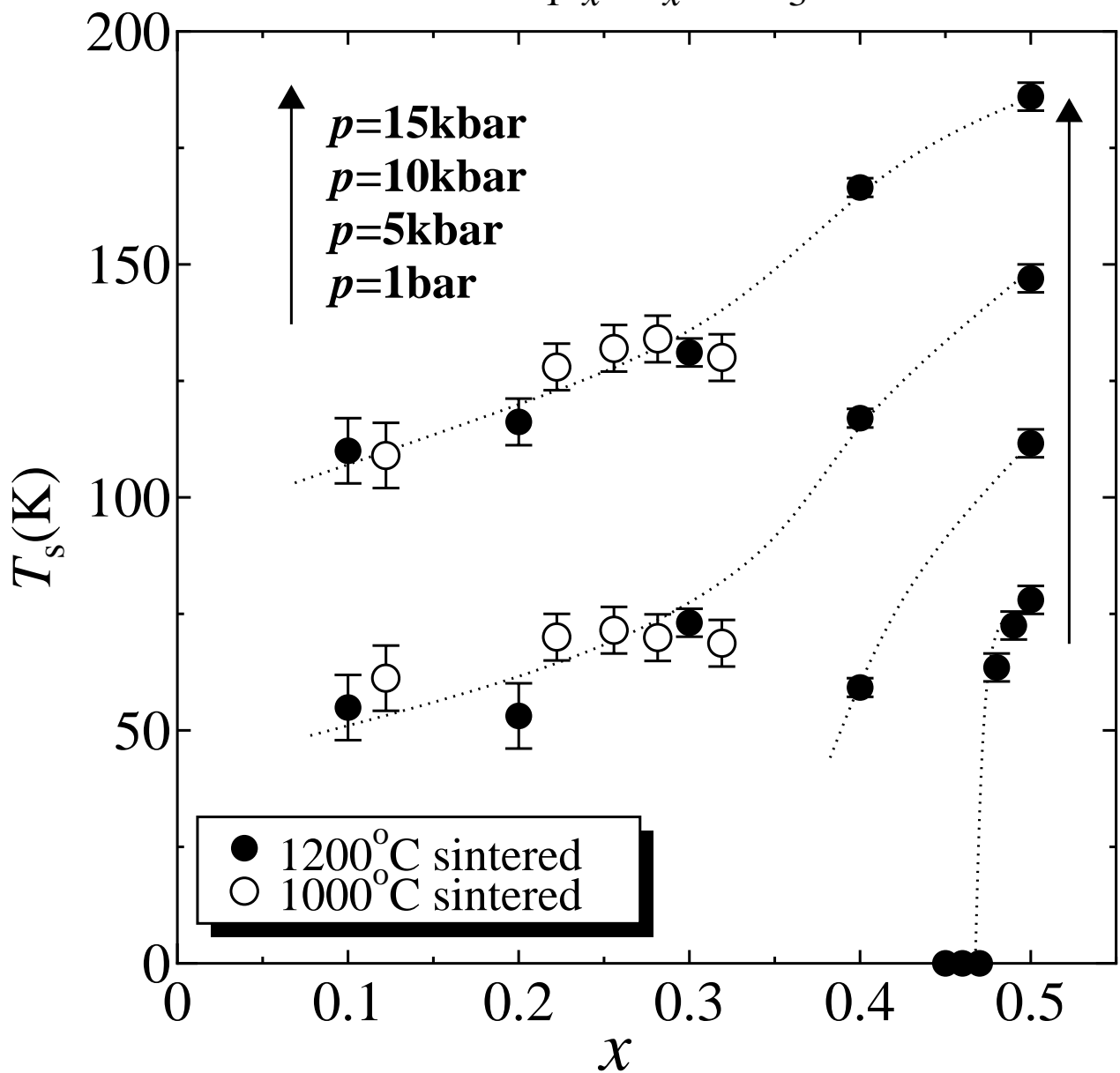

Fig.4
Fujita et al.

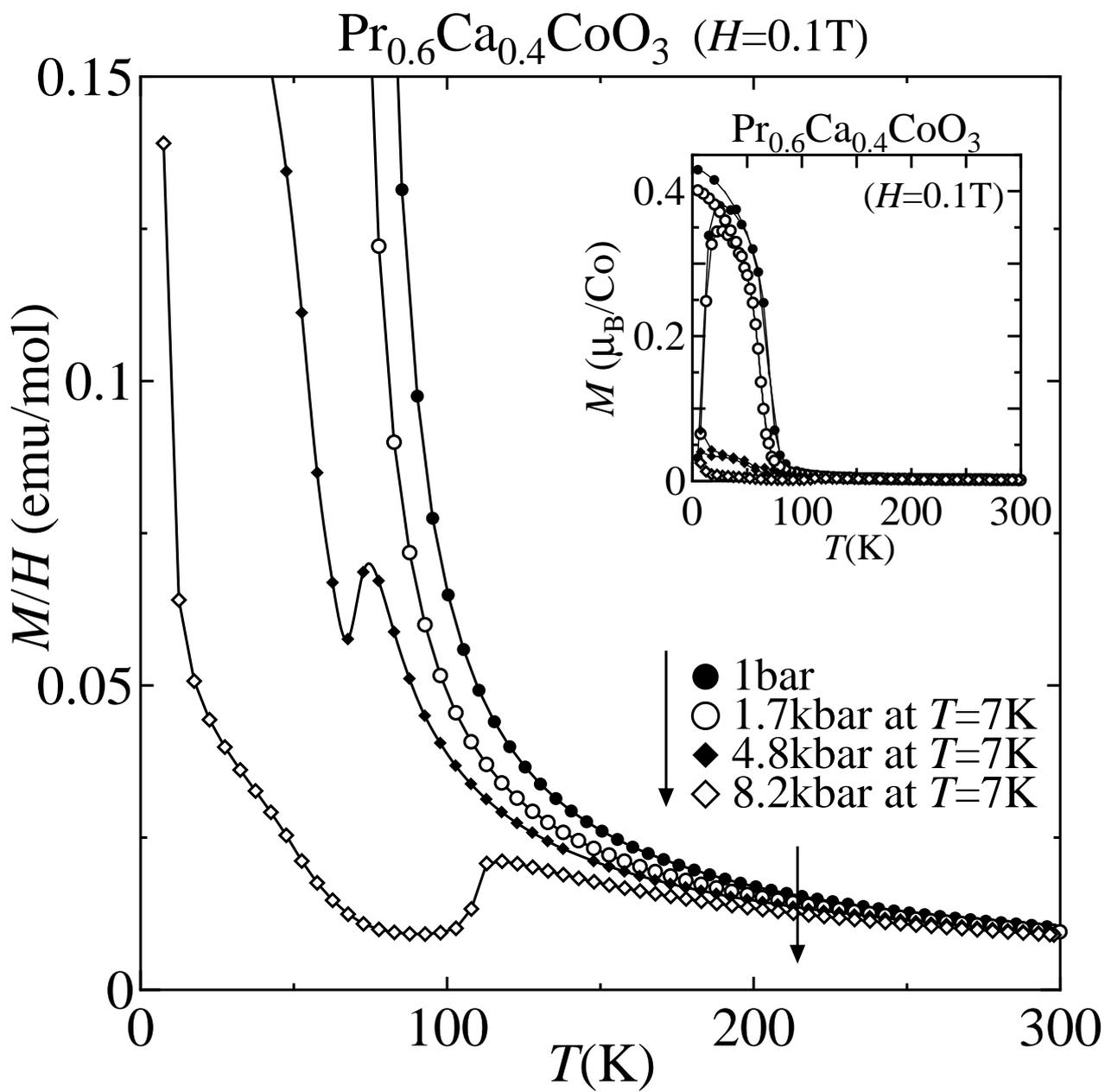

Fig.5
Fujita *et al.*

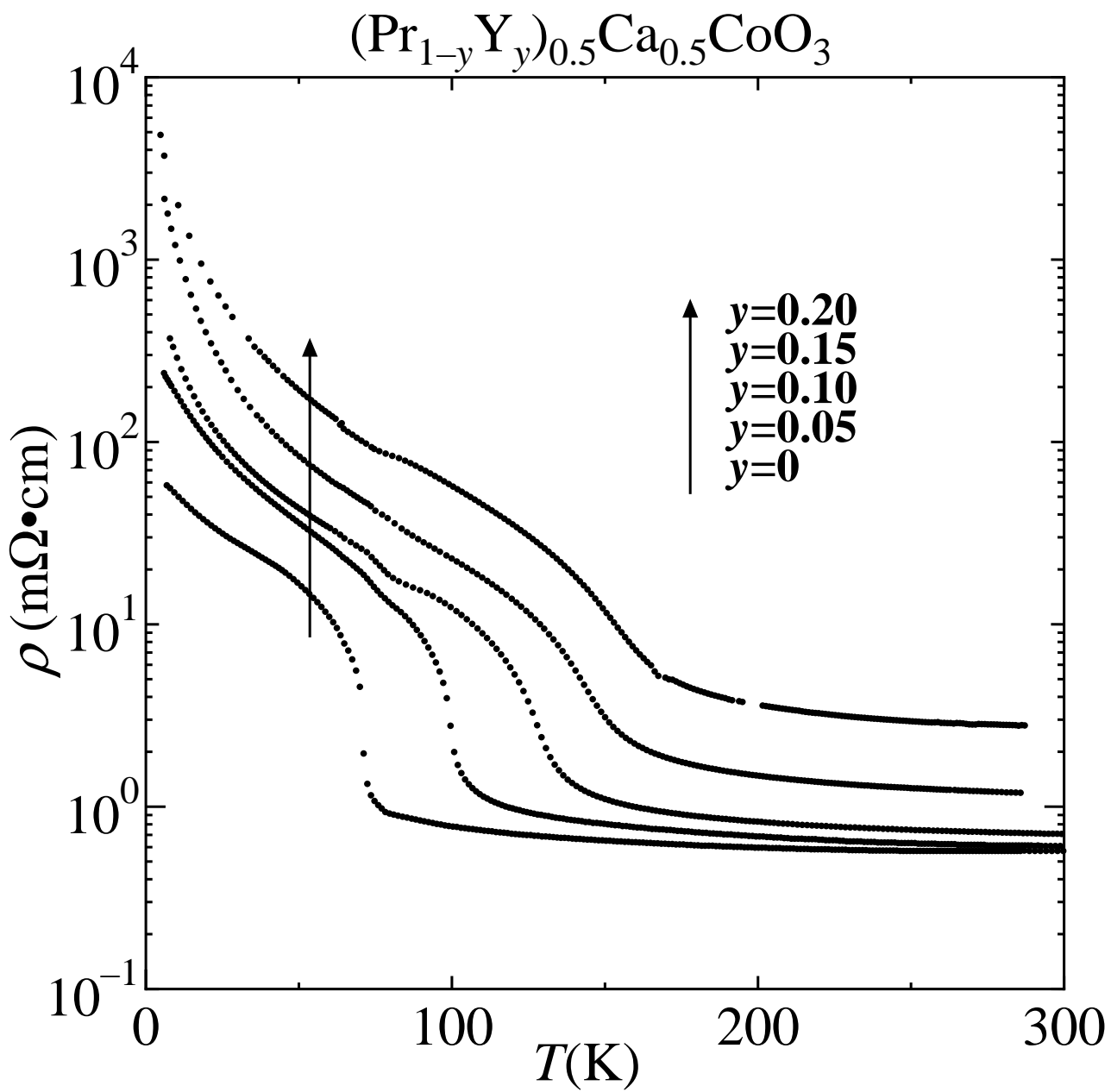

Fig.6(a)

Fujita *et al.*

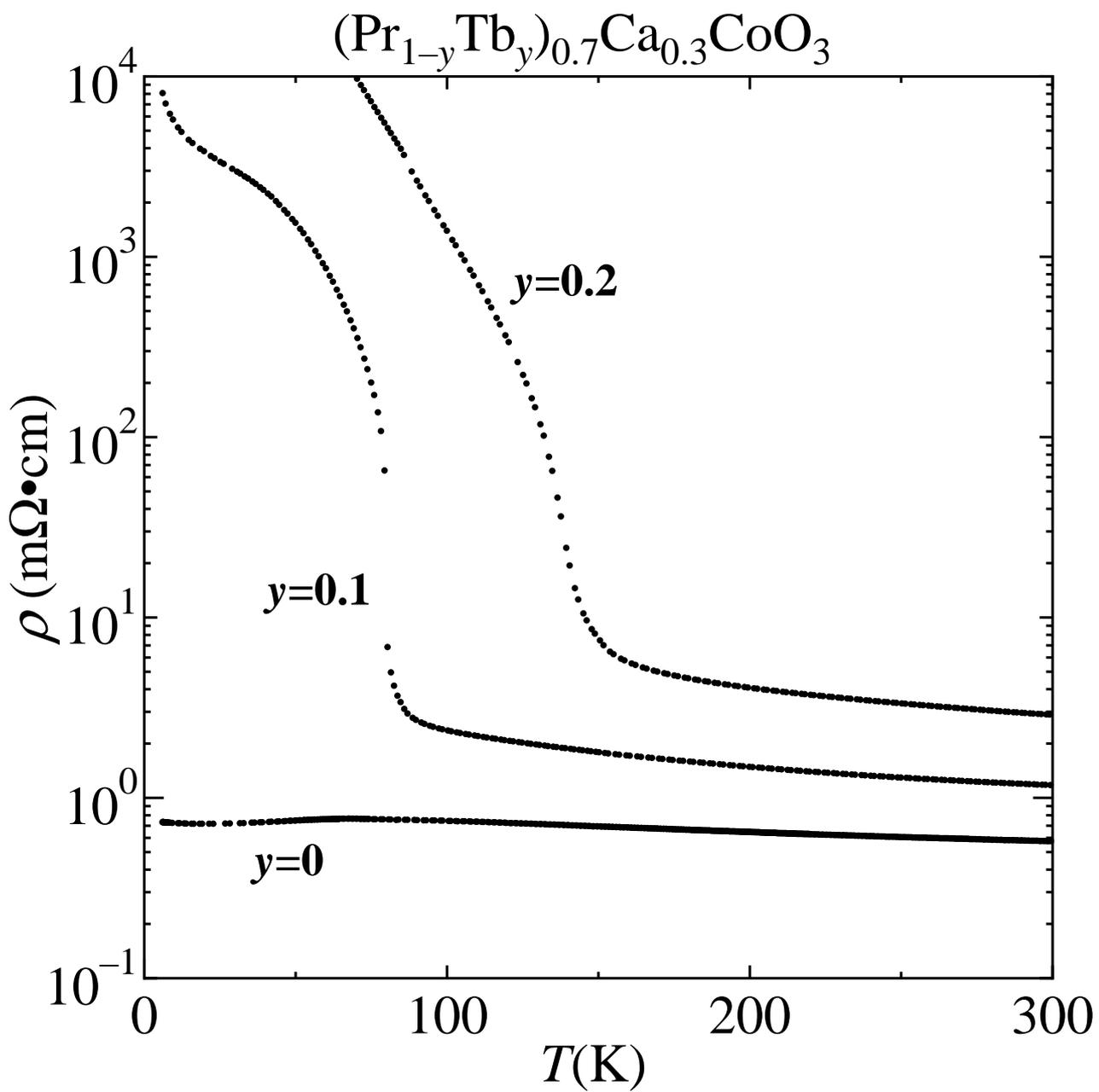

Fig.6(b)
Fujita *et al.*

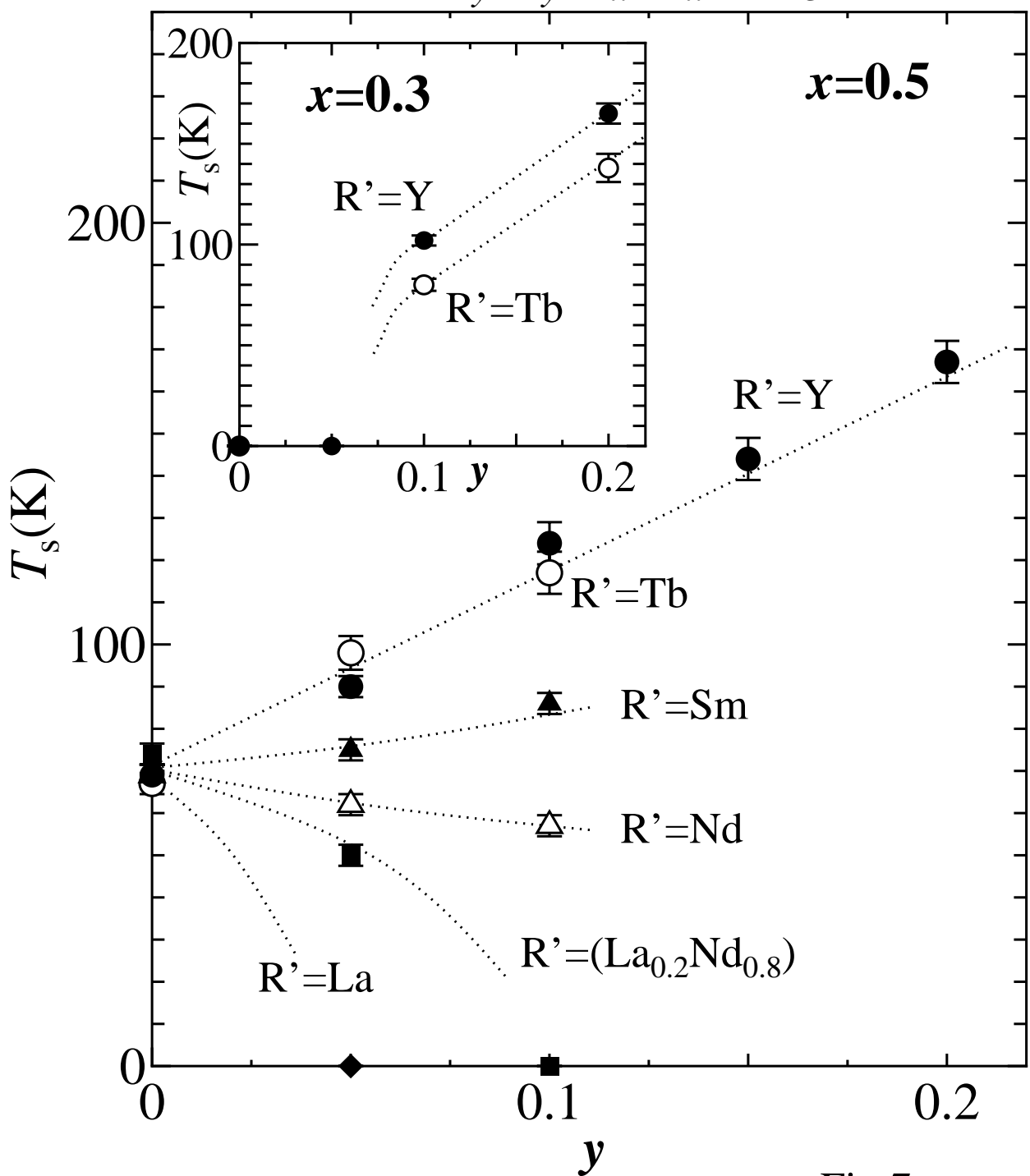

Fig.7
Fujita et al.

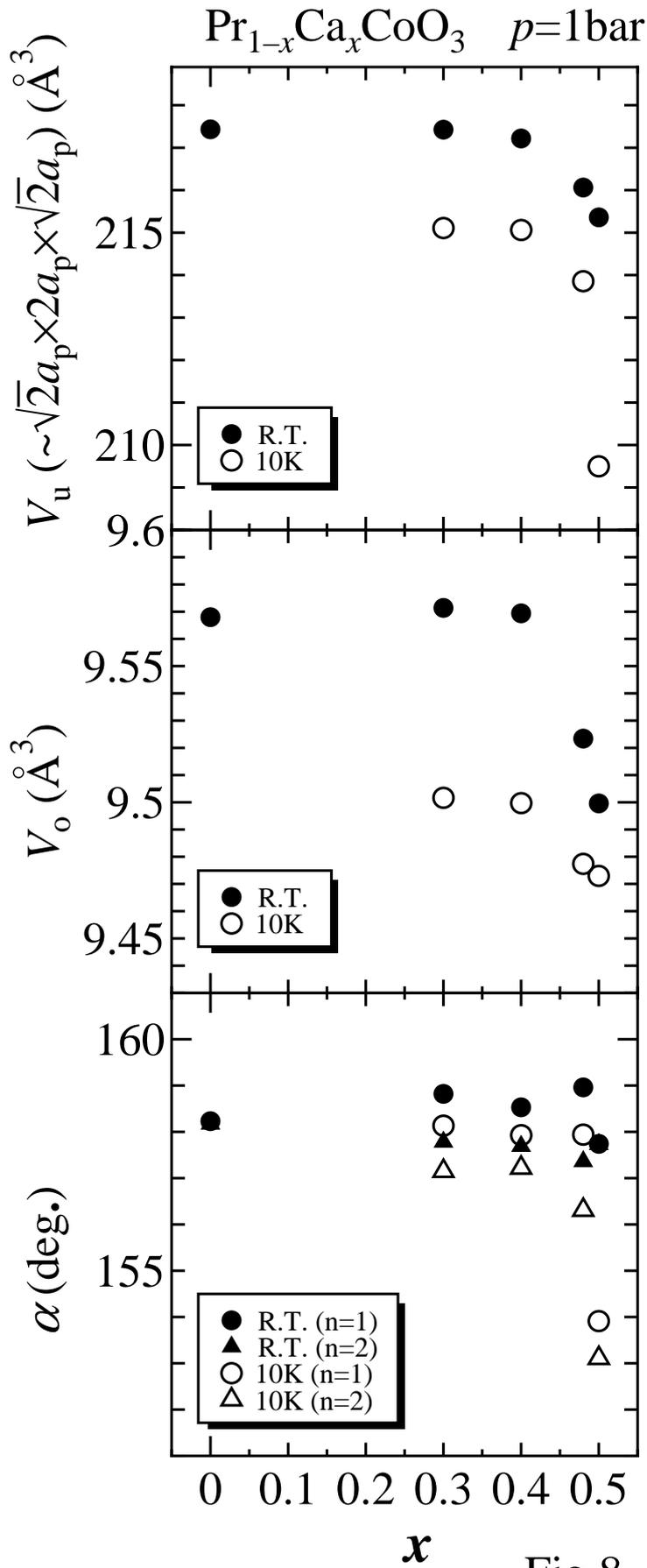

Fig.8
Fujita *et al.*

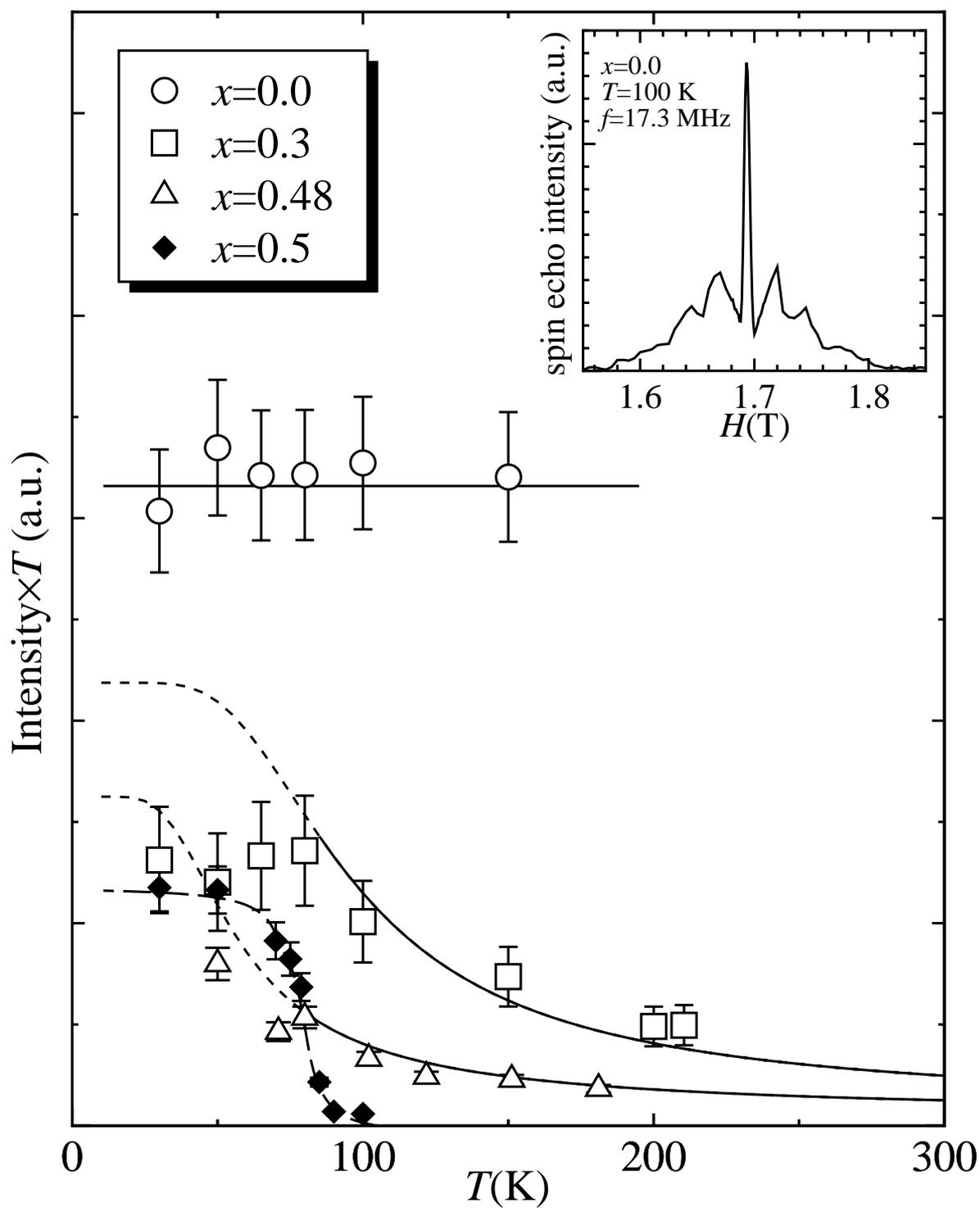

Fig.9

Fujita et al.

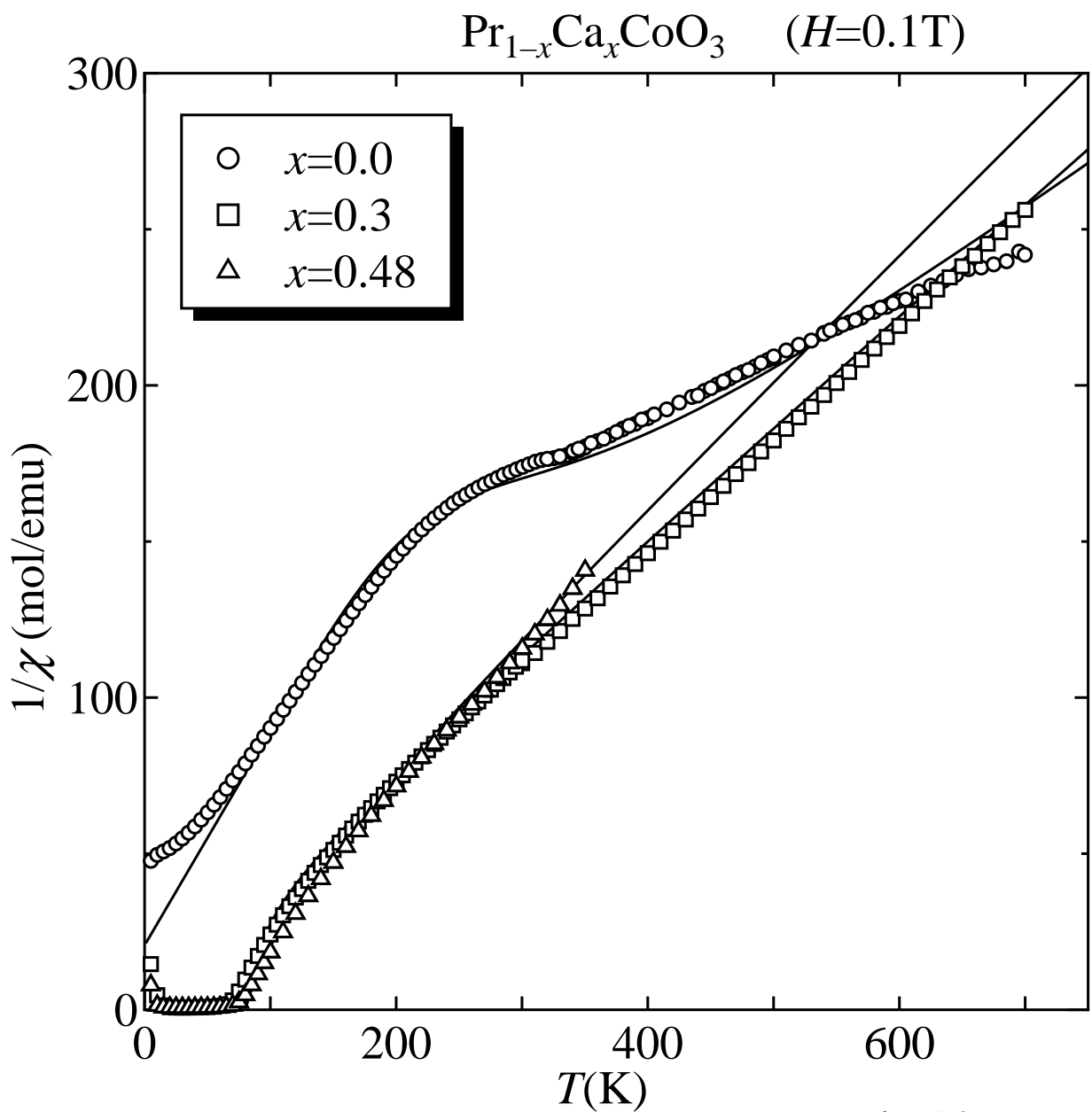

Fig.10
Fujita *et al.*

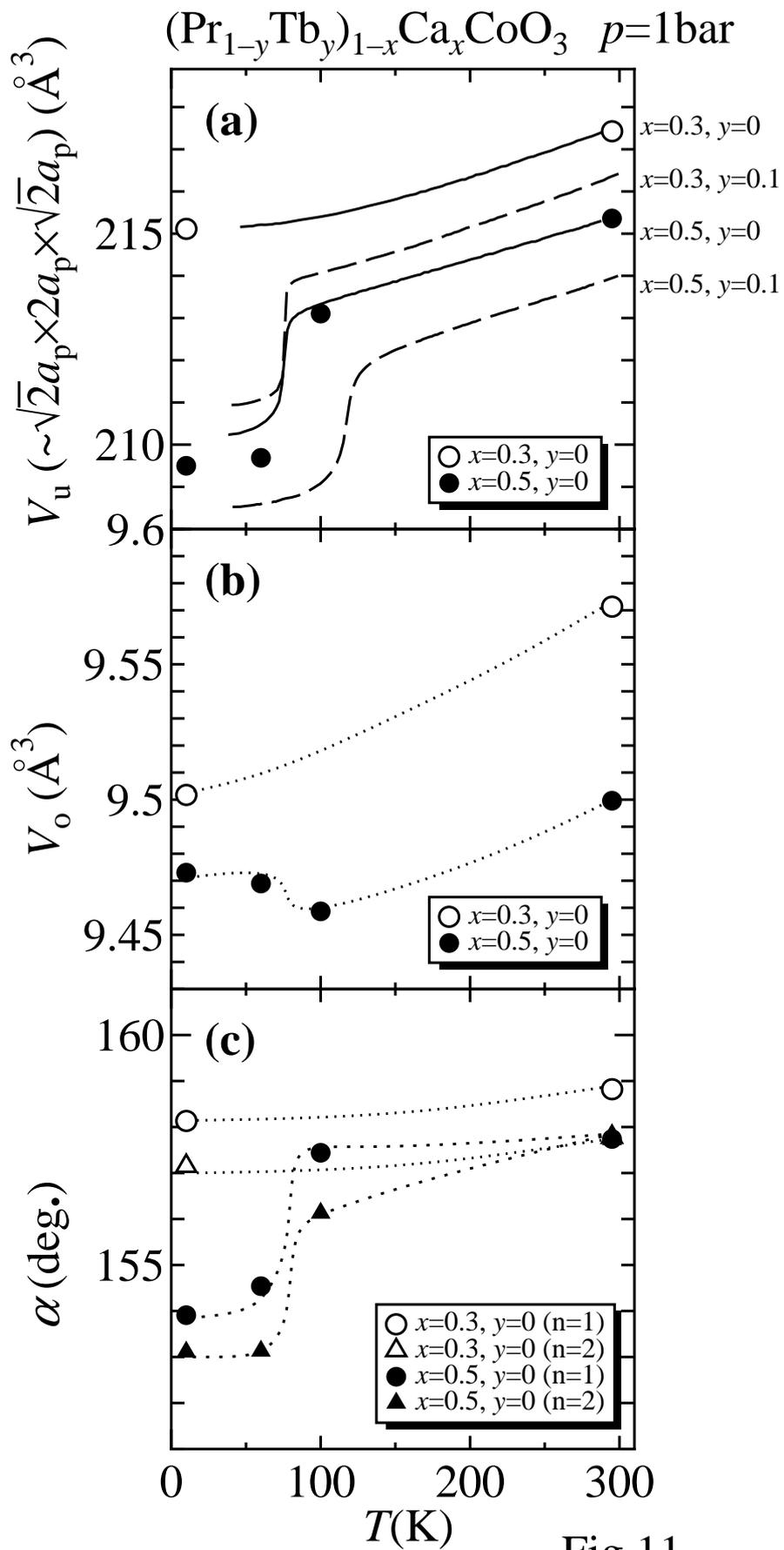

Fig.11 Fujita et al.

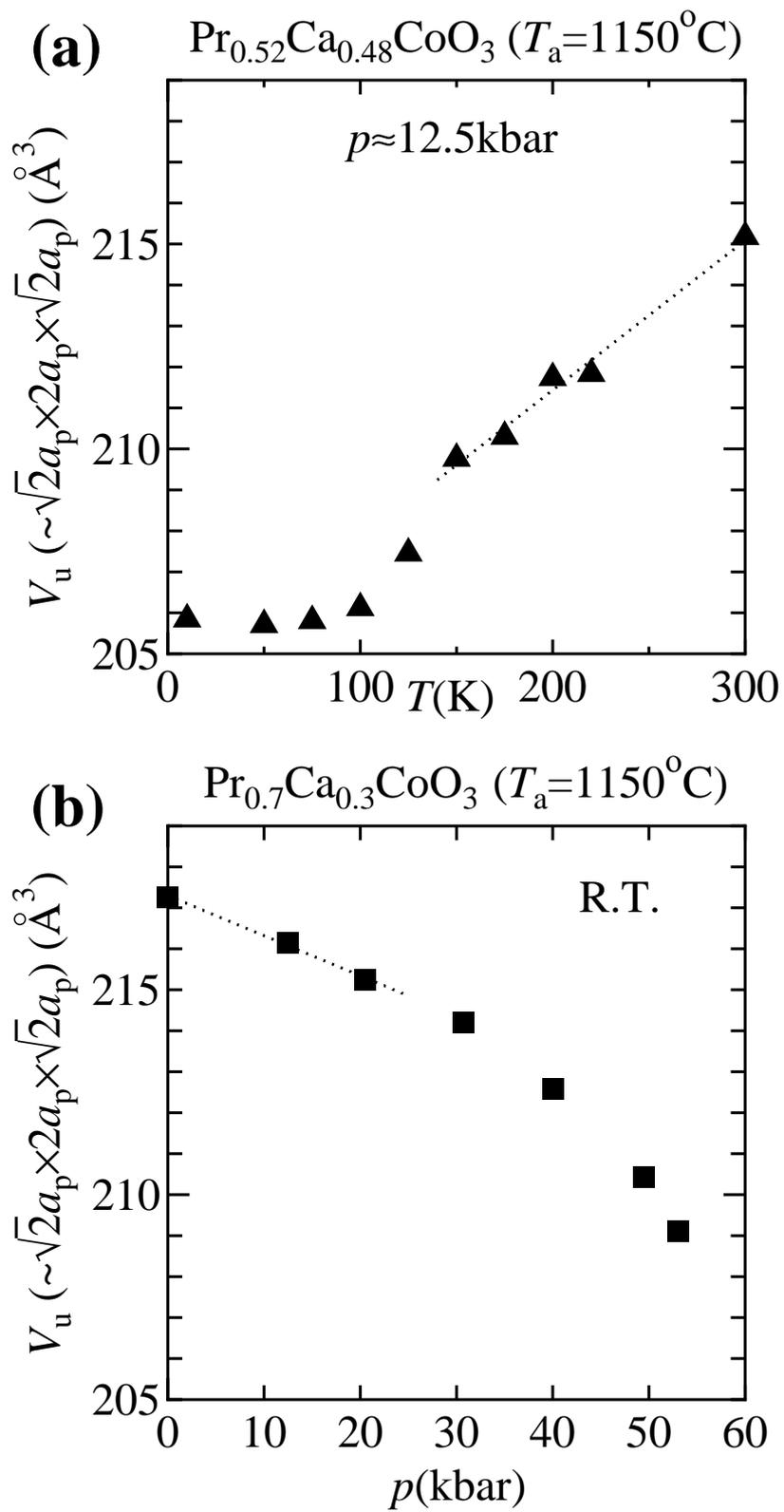

Fig.12
Fujita *et al.*

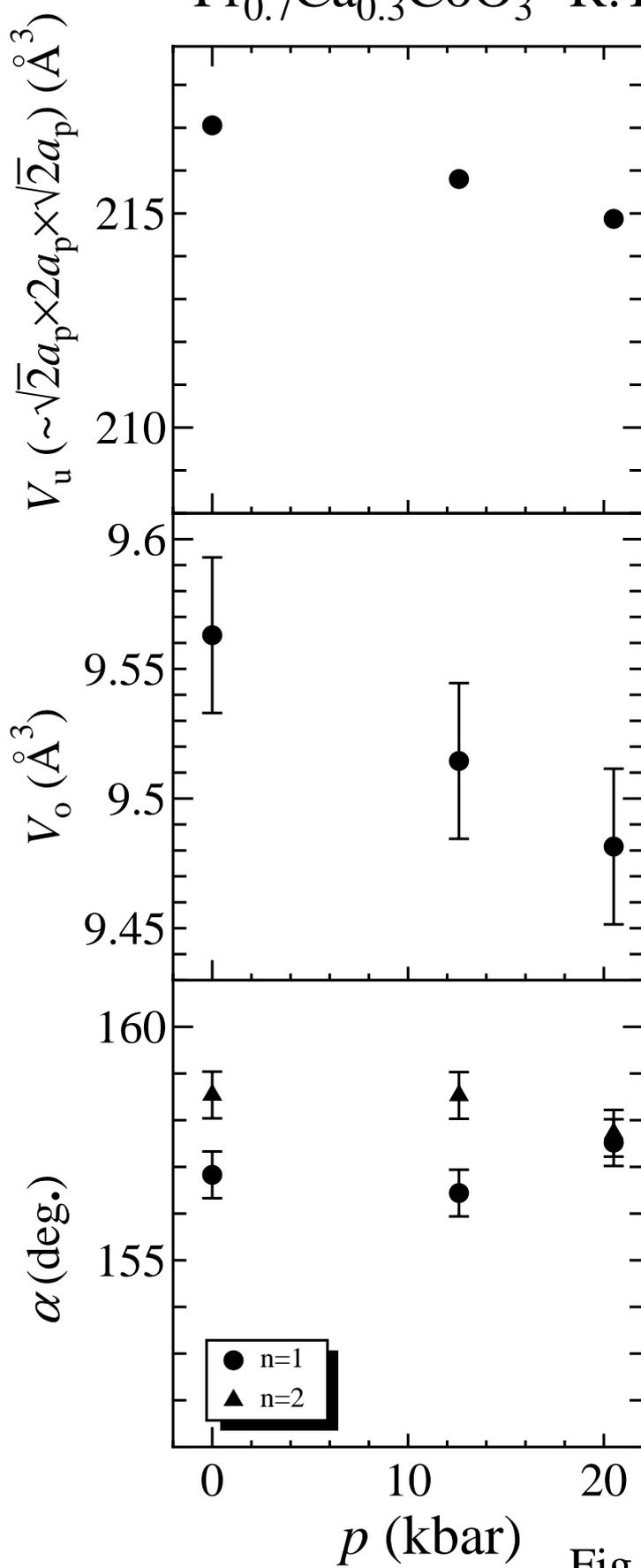

Fig.13
Fujita *et al.*

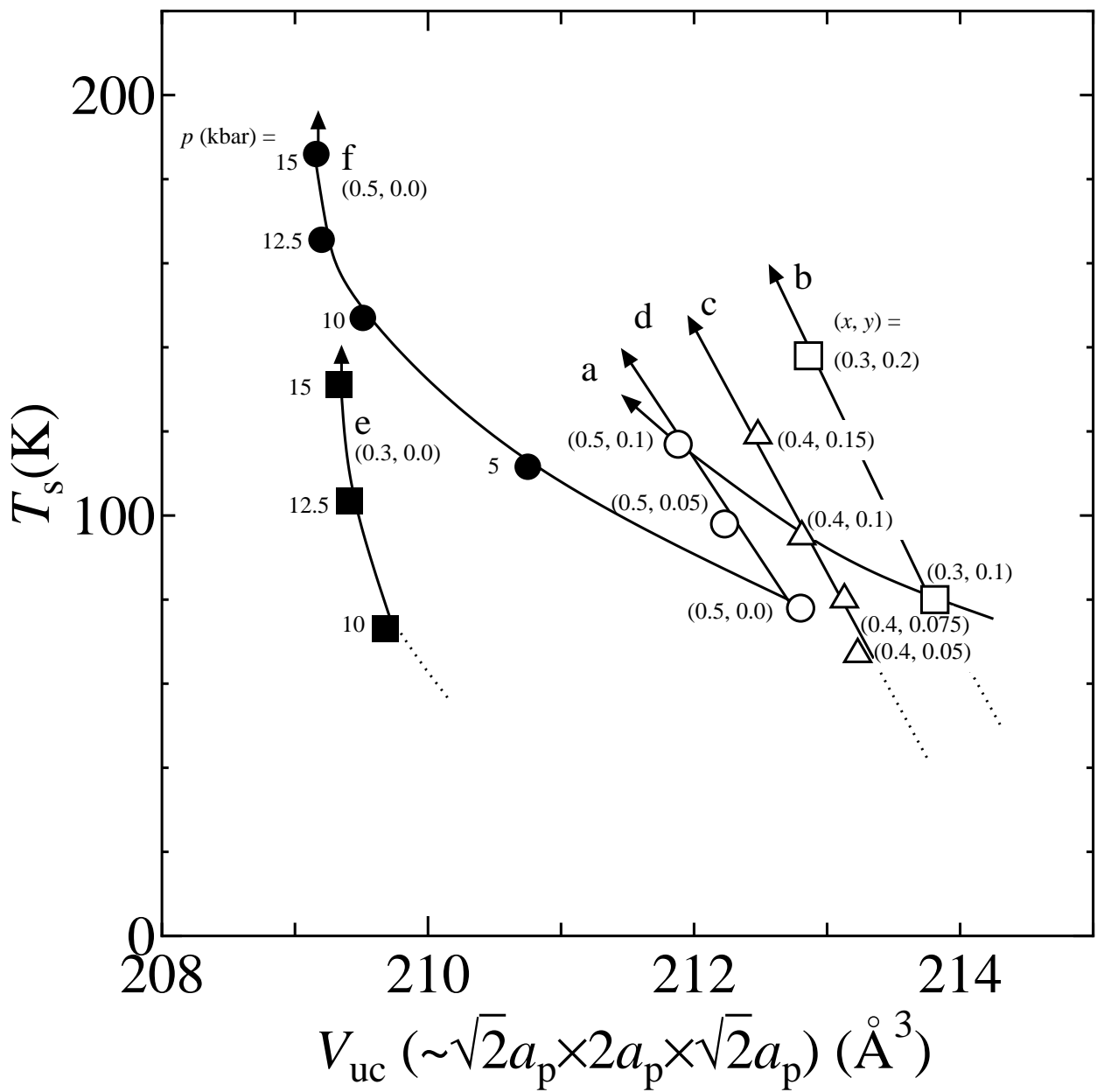

Fig.14 Fujita et al.